\documentclass[iop]{emulateapj}
\usepackage[usenames, dvipsnames]{color}

\shorttitle{VVV Survey Microlensing}
\shortauthors{Navarro et al. 2018}

\begin{document}

\title{VVV Survey Microlensing: the Galactic Longitude Dependence}

\author{Mar\'ia Gabriela Navarro \altaffilmark{1,2} 
Dante Minniti \altaffilmark{1,2,3} 
Rodrigo Contreras Ramos \altaffilmark{2,4}}


\affil{Departamento de Ciencias F\'isicas, Facultad de Ciencias Exactas, Universidad Andr\'es Bello, Av. Fern\'andez Concha 700, Las Condes, Santiago, Chile}
\affil{Millennium Institute of Astrophysics, Av. Vicuna Mackenna 4860, 782-0436, Santiago, Chile}
\affil{Vatican Observatory, V00120 Vatican City State, Italy}
\affil{Instituto de Astrof\'isica, Pontificia Universidad Cat\'olica de Chile, Av. Vicuna Mackenna 4860, 782-0436 Macul, Santiago, Chile}

\begin{abstract}
We completed the search for microlensing events in the zero latitude area of the Galactic bulge using the VVV Survey near-IR data obtained between 2010 and 2015. We have now a total sample of $N=630$ events
Using the near-IR Color-Magnitude Diagram we selected the Red Clump sources to analyze the longitude dependence of microlensing across the central region of the Galactic plane.
The events show a homogeneous distribution, smoothly increasing in numbers towards the Galactic centre, as predicted by different models.
We find a slight asymmetry, with a larger number of events toward negative longitudes than positive longitudes. This asymmetry is seen both in the  complete sample and the subsample of red clump giant sources, and it is possibly related with the inclination of the bar along the line of sight. 
The timescale distribution is fairly symmetric with a peak in $17.4\pm1.0$ days for the complete sample ($N=630$ events), and $20.7\pm1.0$ days for the Red Clump stars ($N=291$ events), in agreement with previous results.
\end{abstract}




\section{Introduction}

The largest surveys dedicated to detect bulge microlensing events so far, such as the Massive Astrophysical Compact Halo Objects (MACHO; \citealt{Alcock93}), the Optical Gravitational Lensing Experiment (OGLE; \citealt{Udalski93}),  the Microlensing Observations in Astrophysics (MOA; \citealt{Bond01}),  the Exp\'erience pour la Recherche d'Objets Sombres (EROS; \citealt{Aubourg93}), the Disk Unseen Objects (DUO; \citealt{Alard95}), the Wise Observatory \citep{Shvartzvald12} and more recently the Korean Microlensing Telescope Network  (KMTNet; \citealt{Kim10}, \citealt{kim17}), have detected tens of thousands of events toward the Galactic bulge. However, the central Galactic plane (the low latitude regions with $|b|<2^\circ$ surrounding the Galactic centre) has not been studied because of the extremely high differential reddening and source crowding.
In addition to being an unexplored area, it is an interesting area to study since the number of microlensing events is expected to increase toward the central Galactic plane region \citep{Gould95, lukaz15, Shvartzvald17}.
Moreover, the complete analysis of microlensing events at low latitudes and longitudes is very useful to optimize the observational campaign for theWide Field Infrared Survey Telescope (WFIRST) (\citealt{Green12}, \citealt{Spergel15}).

With the new era of near-IR surveys, such as the UKIRT microlensing survey (\cite{Shvartzvald17}, \cite{Shvartzvald18}) and the \emph{VISTA Variables in the V\'ia L\'actea Survey} (VVV; \citealt{minniti10}), we are now able to penetrate the gas and dust to study the central Galactic plane in detail. 
In this framework we completed the search in the 14 VVV tiles located in the bulge centered at $b=0^\circ$. The proof of concept and initial results for the innermost three tiles (b332, b333, b334) were published in \cite{navarro17}. We  found 182 microlensing events, with an excess in the number of  events in the central tile b333 in comparison with the other two tiles, and a relatively large number of long timescale events (with $t_E>100$ days). 
In this report we extend the area coverage, based on the 5-years long campaign of the VVV survey near-IR observations, and we present the  spatial and timescale distribution analysis based on the final sample of microlensing events that is  3.5 times larger.

In Section \ref{sec:sec2} the data used to do the search and the method is presented. In Section \ref{sec:sec3} we describe the analysis of the Color-Magnitude diagrams (CMDs) and in Section \ref{sec:sec4} we discuss the spatial distribution of the sample. The timescale analysis is discussed in Section \ref{sec:sec5}. Finally our conclusions are presented in Section \ref{sec:sec6}.

\section{Observations and Method}
\label{sec:sec2}

We use the data from the \emph{VISTA Variables in the V\'ia L\'actea Survey} (VVV; \citealt{minniti10}), that is a near-IR variability Survey that scans the Milky Way Bulge and an adjacent section of the mid-plane using the  \emph{Visible and Infrared Survey Telescope for Astronomy } (VISTA), a 4 m telescope located at ESO's Cerro Paranal Observatory in Chile.
The  PSF photometry was carried out with DAOPHOT following the procedures described in detail by \cite{contreras17}.
The total area analyzed here comprises 14 VVV tiles in the bulge (from b327 to b340) covering the region within $-10.00^\circ \leq l \leq 10.44^\circ$ and $ -0.46^\circ \leq b \leq 0.65^\circ$. 
These data included from 73 ($b340$) to 104 epochs ($b333$) spanning six seasons (2010-2015) of observations. 
These multi-epoch magnitudes in the $K_s$-band comprise a total of about $63 \times 10^6$ light curves for individual point sources. 

The microlensing event selection follows the same procedure explained in \cite{navarro17}, and Navarro et al. (2018, in preparation). Briefly, we fitted the standard microlensing model which assumes that the source and the lens are point like objects \citep{refsdal64}. 
The final sample contains $630$ microlensing events, 182 of which were previously published in \cite{navarro17}.  
Table~\ref{tbl-1} lists the Galactic coordinates of each tile, along with the number of total light-curves analyzed, the number of events found, and the corresponding numbers of red clump (RC) events.
A small number of duplicate events found in overlap regions (N=36), is already accounted for in this total sample.


\section{Characterization of the Microlensing Events}
\label{sec:sec3}

\begin{figure}[t]
\epsscale{1.2}
\plotone{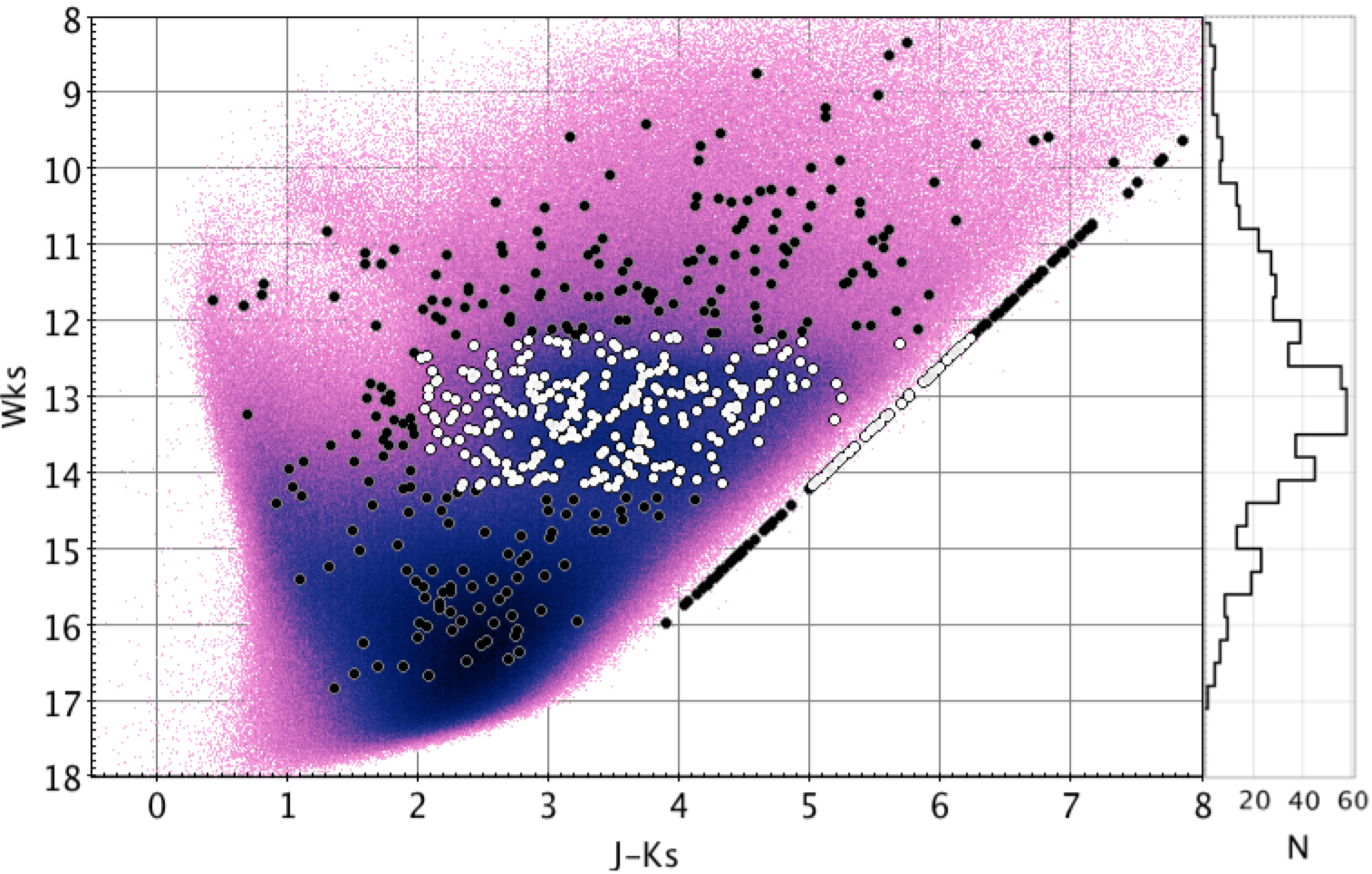}
\caption{Near-IR $W_{K_s}$ $vs$ $J-K_s$ Color-Magnitude diagram for the 14 VVV tiles (from b327 to b340). This is a logarithmic color-coded Hess diagram representation of the 63 million individual sources. The black circles indicate the sources of the sample microlensing events and the white circles are the sources located in the Red Clump. The $W_{K_s}$ histogram of the microlensing events sample is shown in the right panel of the figure. 
\\ \label{cmd_HK}}
\end{figure}

For the microlensing analysis, it is usually better to restrict the sample to the Red Clump Giants stars (RC). These are evolved low mass stars that are burning helium in their cores and are located in a narrow band (Horizontal Branch) of the Color-Magnitude Diagram, hence they act as good distance indicators. 
In the optical microlensing experiments the use of RC stars is preferred because of three main reasons \citep{pop05, sumi05}: 

1. the probability for the source to be located in the bulge is higher,  

2. they are bright and red, so that the photometric completeness is generally higher in the reddened bulge regions, and 

3. they are bright enough that the blending effect might be negligible and therefore, the parameters obtained from the light curve are more reliable. 

The same applies to the near-IR VVV data. To identify better the RC in the CMD of areas where the extinction is extremely high and also variable such as this case, it is more advantageous to use the reddening corrected Wesenheit magnitude, according to the reddening law proposed by \cite{alonso17}, defined as

\begin{equation}
W_k = K_s - 0.428 (J-K_s),
\end{equation}

Figure~\ref{cmd_HK} shows the $W_{K_s}$ $vs$ $J-K_s$  CMD of the area studied in this report plotted as a density map along with the sources of the microlensing events.

Given the low cadence of the VVV survey $K_s$-band observations, it is difficult to separate the microlensing sources from potential blending objects. Therefore, we use the color and magnitude of the baseline sources to identify RC events. We select RC stars using the $W_{K_s}$ histogram shown in the right panel of Figure~\ref{cmd_HK}, according to the clear over density between $12.2 \leq W_{K_s} \leq 14.2$ mag and limiting the color using the CMD in $J-K_s >2$, where $K_s$ is the baseline magnitude and $J$ is the single epoch photometry of the sources. The subsample obtained is composed of $N=291$ events with RC sources. This represents a large fraction ($46\%$) of the total sample.


From the near-IR CMD it is clear that the RC stars are well above the VVV incompleteness limit that sets at about $Ks>17.5$ mag, except for the reddest sources with $J-Ks>5.0$ mag that are affected by the high extinction present in the inner bulge and may be missing from our search, thus we estimate the color using the $J$-mag detection limit of the CMD.
Also, the bluer stars with $J-Ks<1.5$ mag are deemed to be foreground disk sources. We find a total of N=16 disk sources, very few ($\sim 3\%$) in comparison with the total number of events.
The total number of measured stars 
per tile are listed in Table 1, along with the total number of microlensing events and RC events.



Using the extinction ratios for this region from \cite{alonso17}: 
We adopt the mean intrinsic magnitude and color of the RC to be $M_{Ks}= -1.606\pm 0.009$, and $(J-Ks)_0=0.66\pm 0.01$ mag from \cite{ruiz18}.
The blue color cut made here to select RC giants at $J-Ks_0>2.0$, yields extinction and reddening values of $E(J-Ks)>1.34$, and $A_{Ks}>0.57$ mag, respectively.
The reddest RC sources detected in both the J and Ks-bands have $J-Ks=6.0$ implying $E(J-Ks)=5.34$, and $A_{Ks}=2.28$.
Some of the most reddened sources are not detected in the J-band at all, but analyzing the H-band magnitude of the events we note that the reddest RC sources detected in both the H and Ks-bands have $H-Ks\sim 2.3$ yielding $A_{Ks}=2.36$.
Assuming $A_{Ks}/A_V \simeq 0.11$ \citep{Schlafly11}, the reddest sources of microlensing events observed here would have optical extinctions up to $A_V  \sim 21$ mag.
Such microlensing events are beyond detection for current optical surveys, suggesting that our sample can include events with sources in the far disk. 
Only a microlensing search with WFIRST would be capable of improving upon the results from current the near-IR ground based surveys.

\section{Spatial Distribution}
\label{sec:sec4}

\begin{figure*}[t]
  \includegraphics[width=\textwidth]{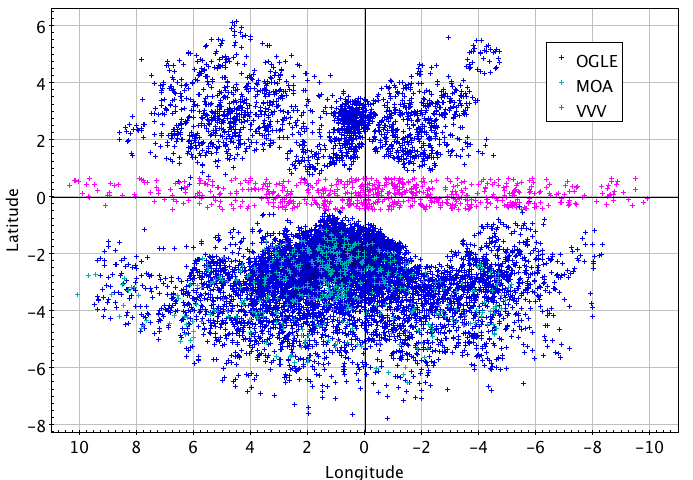}
\caption{Spatial distribution of the new microlensing events (magenta crosses) around the Galactic centre with the events published by OGLE (blue crosses) and MOA (cyan crosses) between the 2010 and 2015. The duplicate events in the overlapping VVV areas have been accounted for. \\  \label{Distribution}}
\end{figure*}

The spatial distribution of the complete sample is shown in Figure~\ref{Distribution}, along with the distribution of events discovered by the OGLE and  MOA optical surveys between years 2010 and 2015 \citep{sumi13, lukaz15}. This figure underscores the importance of the VVV survey that completes the microlensing census at low latitudes, where the optical surveys are blind because of the high extinction.

\begin{figure}
\epsscale{1.2}
\plotone{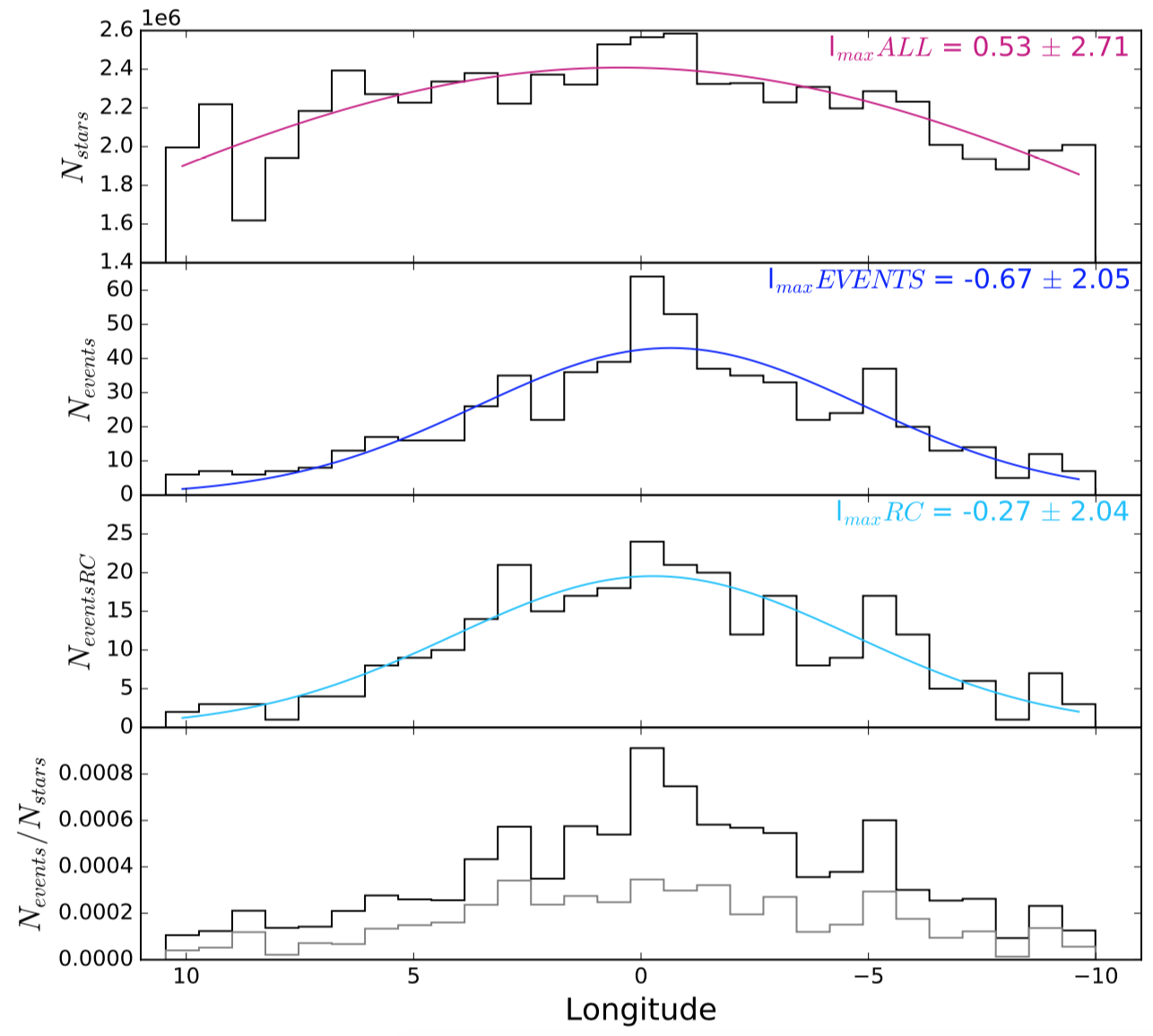}
\caption{Top panel: Histogram of galactic longitude of the $63 \times 10^6$ stars detected in the area within the 14 VVV tiles analyzed.
Second panel: Histogram of galactic longitude of the total events detected in this research.
Third panel: Histogram of galactic longitude of the events located in the Red Clump.
The colored lines are the best Gaussian fit for each case, with the indicated means and sigmas.
Bottom panel: Distribution of the relative number of microlensing events ($N_{events}/N_{stars}$)  as function of Galactic longitude. The upper and lower histograms show the total number of events and RC events, respectively. \\ \label{nevents_ratio}}
\end{figure}

The longitude distribution of the total number of events is shown in Figure~\ref{nevents_ratio}. 
This distribution clearly shows that the number of events increases towards the Galactic centre and confirm the excess of lenses in the Galactic centre found in \cite{navarro17}.
This is not only due to the higher stellar density, but it is a real increase in the microlensing rate.
The number of light-curves analyzed  increases only by about  $20\%$ from the edges ($|l|=10$ deg) to the Galactic centre at $l=0$ deg, while in comparison the total number of events increases by almost a factor of five (c.f. Table~\ref{tbl-1}).
 
Also, the distribution is not symmetric about the Galactic minor axis, as the negative Galactic longitudes exhibit a higher number density of events ($\sim 60 \%$) than the positive ones. 
Additionally there are some inhomogeneities: regions of over-density of events as well as regions with fewer events. 
A visual inspection reveals that the over-densities and under-densities seem to be correlated with reddening, in the sense that the more reddened the field the smaller the number of events. 
If this is the case, the real number of events in the central-most regions that are generally more reddened (like the Galactic centre tile b333) may be larger than observed.


There are two main competing effects to take into account. We expect a higher number of events toward the centre due to the density profile of the galaxy that is known to increase toward the centre. Contrary, the extinction is higher toward the central fields, therefore the number of observed events is expected to become more incomplete as we approach the Galactic centre due to the heavy extinction. 

We use the number of events normalized to the total number of stars per tile. Taking into account the total sample, this fraction is 
$N_{total}=1.0 \times 10^{-5}$.
This is not constant across the plane, as the normalized number of events per number of stars clearly increases towards the Galactic centre.

The use of RC giants should minimize the incompleteness, and we examine the RC events, that is the $46 \%$ of the sample, to strengthen the results obtained from the total sample, namely:

1. There is a smooth increase in the RC event numbers towards the Galactic centre. This was predicted by all the available models.

2. The central number of RC events at $l=0$ deg is $\sim 5$ times the number at $l=10$ deg. 
This effect is more pronounced considering the normalized number of RC events 
$N_{RC events}/N_{stars}$,  where a stronger increase is seen, with the number at the centre being $\sim 8$ times larger than that at $l=10$ degrees. 
The different existing models make different predictions, and this observed number is important to quantitatively fine-tune future models.


3. The RC event distribution is non axisymmetric, with an excess of events at negative longitudes. Excluding the five central tiles with $-3<l<3$, there are $\sim 20 \%$ more RC events in the tiles at negative longitudes with $-10<l<-3$ deg ($N_{RC}=90$ events) than in the tiles at positive longitudes with $  3<l<10$ deg ($N_{RC}=71$ events). 
This result was not predicted by the existing models.
 These results are more evident considering the sampling efficiency corrections (that yield $\sim 32\%$).

This asymmetry in the number of microlensing events at zero latitude can be explained by the inclination of the bar.
At negative longitudes we not only have a longer line of sight before hitting the main bulge sources that are located at the opposite side of the bar, but also the length of the optical path through the bar itself is longer than at positive latitudes. The first effect causes more bulge-disk events, while the latter produces more bulge-bulge events.
Consequently, the probability for detecting microlensing events is larger toward negative longitudes, as observed. However, this detection is of low statistical significance ($\sim 2\sigma$).

Interestingly, the optical microlensing events that map higher Galactic latitudes have not observed such a pronounced asymmetry. This may be due to a strong dependence of the effect of the bar with latitude. The VVV microlensing work allows for the first time to probe these spatial dependences all the way to the Galactic plane and centre, in spite of the strong obscuration in these fields.



\section{Timescale Distribution}
\label{sec:sec5}
The only important physical parameter obtained from the standard microlensing model fitting procedure is the Einstein radius crossing time ($t_E$) also called microlensing timescale. This timescale is related to the mass of the lens but also depends on the relative distances (distance between the observer and the lens $D_L$ and between the observer and the source $D_S$), and the relative transverse velocity \citep{Paczynski86}. Thus, although the Einstein radius crossing time is extremely degenerate, the timescale distribution of the sample gives an indication of the masses and velocities of the lenses. Therefore, the timescale distribution depends on the mass function and the velocity dispersion of the lens population.

 The observed timescale distribution is affected by the detection efficiency, which is discussed in detail by Navarro et al. (2018, in preparation). Briefly, the sampling efficiency is cadence and timescale dependent \citep{mroz17},  therefore was evaluated using Monte Carlo simulations of 10,000 events for each fixed representative timescales (1, 3, 5, 10, 20, 40, 60, 80, 100, 150 and 200 days). This procedure was computed for each VVV tile. \\  For the photometric efficiencies, we used the extensive PSF photometric simulations made for the VVV survey by \cite{valenti16} and \cite{contreras18}. For example, comparing the sample of low amplitude RR Lyrae population with the OGLE catalog yields  a completeness of $90\%$ at $Ks \sim 15$ mag. The observed timescale distribution was corrected accordingly, also excluding the first VVV observation season (2010) because of the small number of observations.

\begin{figure}
\epsscale{1.2}
\plotone{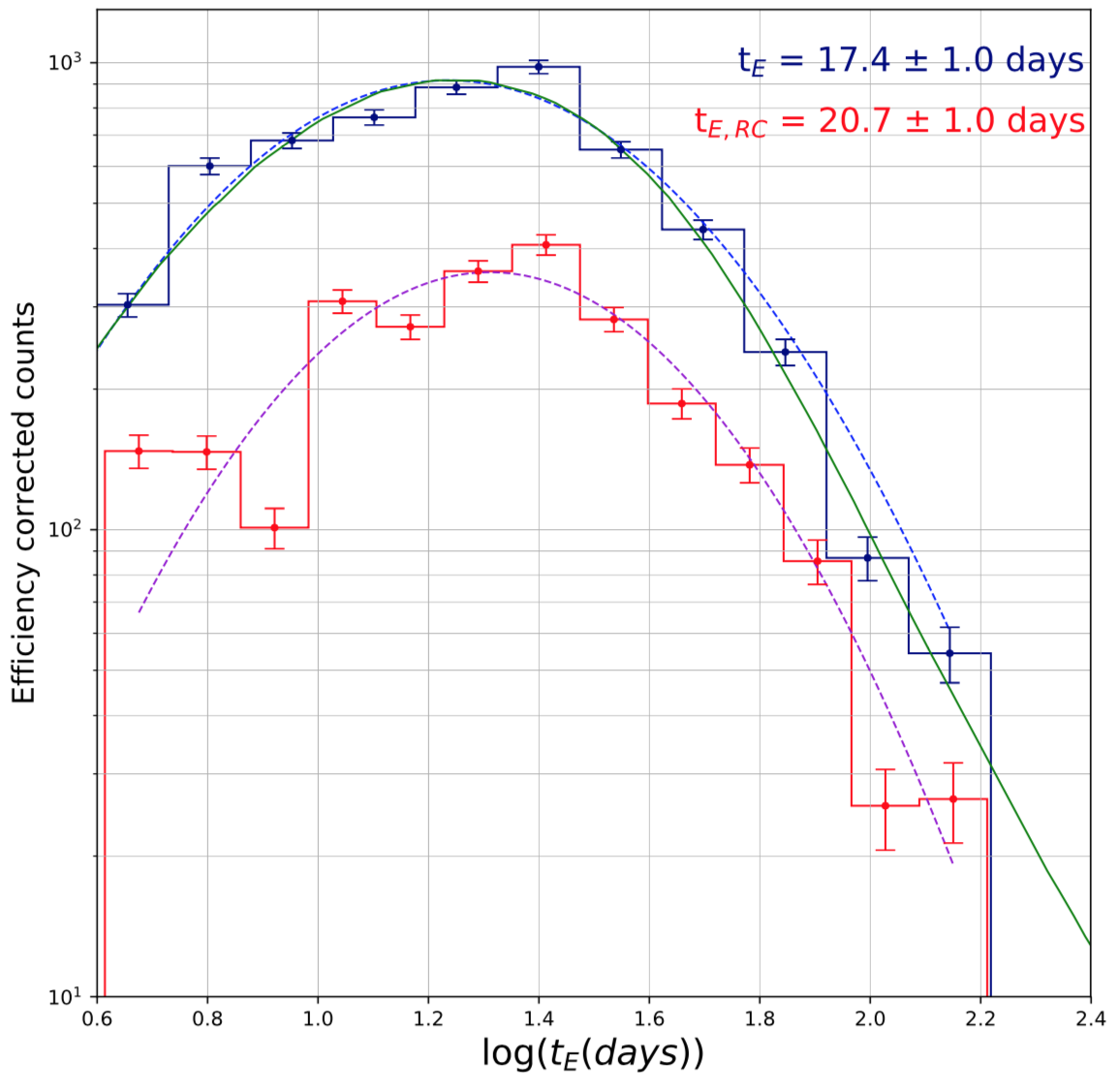}
\caption{ Distribution of the efficiency corrected timescales with blending excluding the first year of VVV observations, for the complete sample (in blue), and the RC events subsample (in red) along with the error bars for each bin. Dotted lines show the best fit model of each distribution.
The green line shows the lognormal distributions of the model proposed by \cite{wegg16}, arbitrarily normalised to the peak. Poisson error bars for each bin are presented. \\ 
\label{timescale_model}}
\end{figure}

 The efficiency corrected timescale distribution (for the model including blending effects) obtained in this work is shown in Figure~\ref{timescale_model}, along with the models of \cite{wegg16}. The mean timescale is $17.4 \pm 1.0$ days for the complete sample, and $20.7 \pm 1.0$ days for the RC sample. 

 Both distributions are in good agreement, and suggest that the typical lenses are lower main-sequence stars and brown dwarfs. For the model without blending the distribution is similar with values slightly lower toward short timescale events and a mean timescale of $13.9 \pm 1.0$ and $16.5 \pm 1.0$ days for the complete and the RC samples, respectively.
 
Previous studies at higher latitudes, like OGLE, obtained $\langle t_E \rangle \sim 32 $ days and $\langle t_E \rangle \sim 28 $ days for uncorrected and efficiency-corrected cases, respectively \citep{sumi05}. Likewise EROS obtained $\langle t_E \rangle \sim 33 $ days  \citep{afonso03}. The distributions obtained by MOA acquired \citep{sumi13}  $\langle t_E \rangle \sim 24 $ days ($\langle t_E \rangle \sim 19 $ days) for the complete sample (RC sources) and OGLE \citep{lukaz15} inferred $\langle t_E \rangle \sim 22, 20, 24 $ days for positive, central and negative longitudes respectively. Some of the studies just mentioned \citep{sumi13, lukaz15} show changes in the timescale distribution with longitude and latitude, specifically a decrease in timescale towards the central area of the Galaxy. Therefore, as it is the first time this area is analyzed, a straight comparison with previous results can not be done. Our Galactic plane fields have mean timescales shorter than the previous studies, as expected from the model predictions of \cite{awiphan16,wood05} where the trend is also evident.






 Thus, we compare our results with models, such as the one proposed by \cite{wegg16}.  
The model is in correct agreement with our results in the central part, where we found the higher number of events, and in the short timescale tail. However, we observe a small excess of long timescale events ($t_E>100$ days), that needs to be confirmed because it is still within the errors.

\section{Conclusions}
\label{sec:sec6}
We have detected $630$ microlensing events within an  area of $20.68$ deg$^2$ around the Galactic centre ($-10.00^\circ \leq l \leq 10.44^\circ$ and $ -0.46^\circ \leq b \leq 0.65^\circ$) using the deep near-IR VVV Survey photometry. 

This is the first time a longitude analysis of the microlensing event population is done across the central Galactic plane at $b=0^\circ$. We found a decrease in the total number of events with increasing Galactic longitude. That was predicted by the models, partly due to the density of stars that increases toward the Galactic centre. 
Also we found a higher concentration of events toward negative longitudes. This trend is observed both for the full sample and for the RC subsample. 
 This can be explained by the inclination of the bar, as the line of sight towards the negative latitudes is longer, increasing the probability of producing microlensing events.

In order to strengthen the results, it is better to restrict the sample to RC stars for three main reasons: higher probability to be located at the bulge, better completeness, and negligible blending effect. 

 The efficiency corrected timescale distribution also is analyzed for all the sample and the RC subsample. The distribution shows a shorter mean timescale than that obtained in previous studies by surveys such as OGLE, MOA, MACHO and EROS. This result is in agreement with previous observational studies that show a decrease in the mean timescale all the way to the center of the Galaxy.

The comparison of our distribution with the existing models shows a correct agreement in spite of a slight inconsistency at the long timescale tail, that needs confirmation with larger samples.\\

 


 The VVV Survey is a powerful tool to detect microlensing events and to study this population at low latitudes where the optical observations are limited. This can be useful to plan the observations of the WFIRST microlensing survey (\citealt{Green12}, \citealt{Spergel15}). If only the total number of lensing events is concerned, we suggest that a WFIRST survey across the Galactic plane covering the whole bulge would be most profitable for microlensing science, especially using a K-band filter as suggested by \cite{Stauffer18}.

\acknowledgments



\begin{deluxetable}{rrrrr}[]
\tablecaption{
Number of VVV Survey Microlensing Events
\label{tbl-1}}
\tablehead{
\colhead{Tile} & \colhead{(l, b)}  & \colhead{$N_{stars}$} & \colhead{$N_{events}$} & \colhead{$N_{RCevents}$}  
}
\startdata
b327	 &  350.737,  0.140  &    4041549  &  19  &  10 \\
b328	 &  352.196,  0.140  &    3810368  &  19  & 7 \\
b329	 &  353.655,  0.140  &    4246204  &  34  & 17  \\
b330	 &  355.114,  0.140  &    4475365  &  60  & 26  \\
b331	 &  356.572,  0.140  &    4533337  &  55  & 25 \\
b332	 &  358.031, 0.140   &    4644865  &  71  & 32  \\
b333	 &  359.490,  0.140  &    5147969 & 119  & 45 \\
b334	 &     0.949,  0.140  &    4840898  &  75  & 36 \\
b335	 &     2.407,  0.140  &    4586590  &  56  & 35  \\
b336	 &     3.866,  0.140  &    4709192  &  42  & 24  \\
b337	 &     5.325,  0.140  &    4490887  &  33  & 17  \\
b338	 &     6.783,  0.140  &    4568856  &  21  & 8 \\
b339	 &     8.242,  0.140  &    4419377  &  13  &  4 \\
b340  &    9.701,  0.140  &    4245168  &  13  & 5  \\
\hline \\
Total	 &                      &   62760625  & 630 &  291 \\
\enddata
\end{deluxetable} 
\bigbreak


\clearpage

\end{document}